\documentstyle[12pt,psfig,rotate]{article}
\begin{document}
\hoffset-1.0cm
\voffset-2.0cm

\newcommand{\be}{\begin{equation}}
\newcommand{\ee}{\end{equation}}
\newcommand{\ba}{\begin{eqnarray}}
\newcommand{\ea}{\end{eqnarray}}
\newcommand{\nn}{\nonumber \\}


\title{ {\bf Chaotic dynamics of SU(2) gauge fields \\
in the presence of static charges}\footnote{Work supported
by DFG and MTA}
}
\author{T.S.Bir\'o
\\
MTA KFKI RMKI Theory Division \\
Budapest, P.O.Box 49, H-1525 Hungary
}

\date{}
\maketitle

\abstract{We have found in numerical simulations that
the chaoticity of the classical hamiltonian
lattice SU(2) gauge field system is reduced in the
presence of static charges at the same total energy.
The transition to non-chaotic behavior is rather
sudden at a critical charge strength.
The equipartition of chromoelectric and chromomagnetic
energy takes place on a time scale essentially faster than the
leading Lyapunov exponent.
}


\vspace{1.5cm}
\section{Introduction}
\vspace{0.5cm}

Knowledge of the mechanisms responsible for local
equipartition of energy carried by non-abelian gauge fields
and in particular knowledge of the speed
by which the allowed phase space of possible field configurations
is filled are important for the understanding
of processes which lead to equilibrium in the very early
universe and in relativistic heavy ion collisions.
Prime examples for such processes
are baryogenesis during the electroweak phase transition,
the creation of primordial fluctuations in the density of
galaxies, and the thermalization of excited quark matter
in relativistic heavy ion collisions.
Deterministic chaos of extended hamiltonian systems, like
lattice models of non-abelian gauge theory are, is a universal
mechanism of ergodization due to soft classical fields.
Its main effect, the heating up of the system, is in addition to
hard parton scattering.

\vspace{0.5cm}
In the recent past numerical studies of the real-time
evolution of classical gauge field systems demonstrated
chaotic behavior of randomly chosen configurations
with chromomagnetic energy\cite{IJMPC5}.
The maximal Lyapunov exponent (measured in lattice units)
has been found to scale linearly with the energy of the
system (also in appropriate units) making thus an extrapolation
to the infinite dimensional field theory in the classical
continuum limit possible. The linear scaling revealed
that the maximum Lyapunov exponent survives in the
continuum limit, so the real physical system has a finite
entropy density generation rate. Abelian systems on the other hand
do not scale linearly leading to vanishing entropy generation
in the continuum limit. Also analytic studies of simplified
subsystems, like the $xy$ - model helped to understand
the trajectory - defocusing property of the potential energy,
i.e. the magnetic part of the gauge field
interaction\cite{BOOK}.

\vspace{0.5cm}
In order to understand the intriguing coincidence between
the numerically found leading Lyapunov exponent and the gluon
damping rate calculated in high-temperature perturbative
gauge theory one has to assume that the energy is fully
thermalized and the gauge field system evolves to a weakly
interacting gas of quanta, the gluons.
Some arguments in favor of a special physical mechanism
based on color non-diagonal interaction between gluons behind
this coincidence has been discussed in \cite{PRD,PLB}.

\vspace{0.5cm}
We expect that the non-abelian gauge field system is
in its magnetic sector the most chaotic.
In realistic situations, in particular in ultrarelativistic
heavy ion collisions there are, however, matter fields: mainly
quarks but also clusters of several quarks representing a
higher color charge. For some specific signals, expected
to come from such reactions if quark - gluon matter has been
formed, there are heavy color charges which generate
strong chromoelectric fields in string - like configurations.
As a matter of fact the formation and physical behavior
of such strings play also a primary role in understanding
the quark confinement mechanism; the very basis of the existence
of hadrons.

\vspace{0.5cm}
It is therefore of genuine interest to investigate the effect
of such chromoelectric strings spanned between static charges
on the chaotic dynamics of random chromomagnetic gauge field
configurations. Although speculative, it also cannot be excluded
that we learn about the confinement mechanism by simulating
the quantum mechanical ground state of QCD, i.e. the hadronic
vacuum, with random classical chromomagnetic fields.
Indeed a pair of in total neutral color charges, quantized or
classical, connected by a single flux line can only be represented
by a so called Wilson line, the object on which static
quark confinement has ever been studied in four dimensional
euclidean lattice gauge theory. A transition in the chaotic
dynamical behavior of the classical gauge field system
depending on the strength of this flux line, on the string constant,
may therefore also be related to the confinement - deconfinement
phase transition.

\vspace{0.5cm}
Even more speculative but still not totally negligible the idea
that random classical field configurations numerically simulate
quantum dynamics. In fact interference patterns similar to
those observed in basic quantum mechanical two slit experiments
can be created numerically by random discrete maps used in
certain chaos games\cite{NA1,NA2,GAME}. Whether the simulation reported
in the present article has anything to do with such phenomena
must be a subject of future considerations.

\vspace{0.5cm}
We organize this paper as follows. First the basic formuli for
the classical hamiltonian SU(2) lattice gauge field theory are
presented. Then the speed of energy sharing between chromomagnetic
and chromoelectric degrees of freedom with and without static
charges is discussed. Finally the effect of strong flux line
initialization on the leading Lyapunov exponent of the chaotic
gauge field system is described.


\vspace{1.0cm}
\section{The lattice SU(2) model}
\vspace{0.5cm}

In the hamiltonian lattice formulation of classical
SU(2) gauge field theory, which we use to study
its chaotic dynamics, the basic variables are group
valued on each link of a $N^3$ cubic lattice\cite{WILSON,CREUTZ}
\be
U_{x,i} = \exp \left( - \frac{i}{2} \tau^a  ga A_i^a(x) \right) .
\ee
In the SU(2) model, the smallest nonabelian gauge group, the
$\tau^a$-s are the Pauli matrices and the $A_i^a(x)$-s are
corresponding components of the nonabelian vector potential.
The subscript $(x,i)$ identifies a general lattice position
$x$ in the interval $(0, N^3-1)$ and $i$ the spatial direction
of a link $i=0,1,2$. Since we use periodic lattices this link variable
representation is complete; the periodic boundary conditions are
taken into account in an index table of neighboring sites in
each of the 6 possible spatial directions.
The parameters $g$ and $a$ are the bare coupling constant and
the lattice spacing, respectively.
In the classical time evolution actually all
space- or timelike quantities can be measured in units of $a$
and all types of physical energies in units of $g^2a$.

\vspace{0.5cm}
While the vector potential is related to the basic link variable
$U_{x,i}$ the magnetic field strength can be reconstructed
from a product of $U$ matrices along an elementary closed line,
the plaquette:
\be
U_P = U_{x,ij} = U_{x,i} \cdot U_{x+i,j} \cdot U_{x+i+j, -i}
\cdot U_{x+j, -i}.
\ee
In the continuum limit $a \rightarrow 0$ the subleading term is
proportional to the chromomagnetic field
\be
U_{x,ij} = \exp \left( -\frac{i}{2}\tau^a g^2a \epsilon_{ijk}
B^a_{x,k} \right).
\ee
Here $\epsilon_{ijk}$ is the totally antisymmetric three dimensional
tensor. In practical calculations it is advantageous to use
the complement link variable, $V_{x,i}$, instead of the plaquette
matrix $U_P$. It is defined so, that the sum of the four
plaquettes leaning on a given link gives exactly
\hbox{$U\cdot V^{\dag}$}.

\vspace{0.5cm}
The chromoelectric field is related to the time derivative of the
basic link variable, $\dot{U},$
\be
E^a_{x,i} = -\frac{ia}{g^2} {\rm tr} \left( \tau^a
\dot{U}_{x,i} U^{\dag}_{x,i} \right) .
\ee
An exactly good albeit not equivalent definition is
\be
E^a_{x,i} = -\frac{ia}{g^2} {\rm tr} \left( \tau^a
U^{\dag}_{x,i} \dot{U}_{x,i} \right) .
\ee
Again, in practical simulations it is more advantageous to use
the canonical momentum $P = \dot{U}$ attached to each link
instead. The Hamiltonian of the SU(2) lattice gauge theory
is given by
\be
H = \sum \left( \frac{1}{2} \left< P,P \right> \,\, + \,\,
1 - \left< U,V \right> \right)
\ee
in terms of link variables. The summation runs over all links.
Here the scalar product of two group elements is given by
\be
\left< A,B \right> = \frac{1}{2} {\rm tr} (AB^{\dag})
\ee
in the matrix representation and by the usual scalar product
of two real four dimensional vectors in the quaternion
representation.
The dynamics generated by this Hamiltonian conserves the
unitarity of the group elements \hbox{det$U=<U,U>=1$}, the orthogonality
of the basic variables and the corresponding canonical conjugates
$\left< P,U \right> = 0$ , and commutes with the classical
Gauss' law. The latter describes the conservation of color
charge locally, i.e. conserves the following quantity calculated
on each lattice sites:
\be
\Gamma = \sum_+  PU^{\dag}  - \sum_- U^{\dag}P,
\label{GAUSS}
\ee
where the "+" summation runs over links originating in the
site and the "-" over those links which end at the site.

\vspace{0.5cm}
For the numerical solution of the equations of motion we
applied a mixed explicit - implicit algorithm which
exactly fulfills the constraint of the Noether charge conservation
also for discrete timesteps. A derivation of this property
is given in \cite{ALGO}.
The implicit algorithm,
\ba
U' &=& U + (P' - \epsilon U) \nn \nn
P' &=& P + (V - \mu U + \epsilon P' )
\label{IMP}
\ea
with
\be
\epsilon = \left< U,P' \right> \qquad {\rm and} \qquad
\mu = \left< V,U \right> + \left< P',P' \right>,
\ee
where the primed quantities denote the updated (new) values
and the elementary timestep is included in the usage of $P= dt \dot{U}$,
can be resolved into explicit equations. Eventually it can be comprised
into the following algorithm
\ba
V_T = V - \left< U, V \right> \, U  &\qquad \qquad &
\tilde{P} = P + V_T \nn \nn
c &=& \sqrt{ 1 + \left< \tilde{P},\tilde{P} \right> } \nn \nn
P' = ( \tilde{P} + U ) / c - c U &\qquad \qquad &
U' = c U + P'.
\ea
It is important to recalculate implicit recursion formuli
analytically, because explicit equations are not only faster
on computer but they also conserve the Noether charge
(in our case Gauss' law) exactly, i.e. up to the highest precision
used in the computer code independently of the discrete time step.
An implicit algorithm, like (\ref{IMP}), would have been solved
iteratively so the exactness of the charge conservation depended
on the iteration depth.

\vspace{0.5cm}
The explicit numerical algorithm we use conserves Gauss' law, i.e.
the lattice covariant divergence of the nonabelian electric field
(eq.\ref{GAUSS}). The only remaining task is to fulfill it
initially. For a purely chromomagnetic configuration, investigated
dynamically up to now, it is trivial to set the color charge
$\Gamma$ on each lattice site to zero by initializing $P=0$
on all links. This state can also be viewed as a
"chromoelectric vacuum".

\vspace{0.5cm}
To describe localized, i.e. on the lattice
pointlike, color charges so that the total configuration is
color neutral, is far from trivial due to the nonabelian
nature of Gauss' law in the SU(2) gauge theory.
In fact no general nonperturbative solution of the classical
Gauss' law is known for arbitrary color charge density.
We follow here, however, a simpler formulation of the
problem: how to initialize
the variables $P$ on the links once a $U$ - configuration is
given, the total system is color neutral and the charges are
localized. Our construction principle is as follows.

\begin{enumerate}

\item   We generate a $U$ - configuration randomly
        (microcanonically) which represents purely chromomagnetic
        energy.

\item   We prescribe a flux line by a sequence of $3N$ neighboring
        lattice sites. In particular we choose a diagonal staircase
        string across the $N^3$ cube.

\item   On the first link of this flux line
        $ P_1 = Q U_1 $ is initialized.

\item   In order not to have any charge in between we initialize
        \newline \hbox{$P_n = U^{\dag}_{n-1} P_{n-1} U_n$}
        for $(1 < n < 3N)$ recursively along the prescribed flux line.

\item   Arriving at the end of the line the last site has a single
        incoming link with nonzero $P$ on it. Its uncompensated
        charge is \hbox{$\Gamma = - F^{\dag} Q F$}
        where $F$ is the path ordered product of basic $U$ variables
        along the flux line (a Wilson line):
        \be F = \prod_{n} U_n. \ee

\item   Finally in order to ensure color neutrality of the system
        we need \hbox{$Q - F^{\dag}QF = 0$} for the total charge
        and \hbox{${\rm tr} \,\, Q = 0$} for ensuring orthogonality of
        $P$ and $U$ everywhere. The only solution to this problem
        is \be Q = \frac{q}{2} \left( F^{\dag} - F \right), \ee
        where the real parameter $q$ describes the strength of
        the flux line or equivalently the magnitude of the end
        charges.

\end{enumerate}


\vspace{1.5cm}
\section{Less chaos due to flux lines}
\vspace{0.5cm}

We measured the leading Lyapunov exponent as the
logarithmic time-derivative of the gauge invariant distance,
$D(t),$ between two initially adjacent field configurations
\be
h = \frac{d}{dt}  \ln D(t).
\ee
This coincides with the original definition of the Lyapunov
exponent whenever $\ln D(t)$ linearly grows.
The numerical value of $h$ was fitted in such time-intervals
only using a time step $a=0.005$ equal to the lattice spacing for
the sake of simplicity. The gauge invariant distance of
two field configurations is defined by
\be
D(t) = \sum  \left| \left< U(t), V(t) \right>
\,\, - \,\, \left< U^*(t), V^*(t) \right> \right|,
\ee
where the superscript $^*$ denotes a configuration only
slightly different from the unstared one initially and
the summation runs over all links.
This definition of gauge invariant distance differs from
the one used earlier \cite{IJMPC5}  - defined using the
plaquette sum - only in the actual value of the distance
but not in its growth rate from which we obtain the
leading Lyapunov exponent.

\vspace{0.5cm}
The initially small distance between two trial configurations
we produce by rotating randomly all link quaternions, $U$,
by an SU(2) group element near to unity.
Although -- in principle -- we would have to average the
value of the leading exponent obtained this way over the
phase space, the ergodizing property of the chaotic evolution
ensures that both trajectories sample almost the whole
available phase space in a single long term run.

\vspace{0.5cm}
\begin{figure}[p]
\vspace{0.3cm}
\centerline{\rotate[r]{\psfig{figure=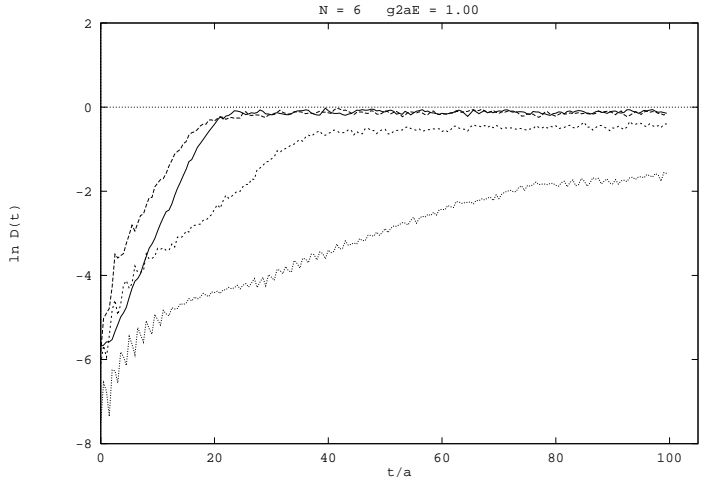,width=6.0cm}}}
\vspace{0.3cm}
\caption{\label{MDFIG} The evolution of the gauge invariant distance
between two initially adjacent configurations with
different static charges at the ends of a diagonal
flux line in a $6^3$ cubic lattice. The total energy
per link per color is $g^2aE=1.00$ for all curves.}
\end{figure}

\vspace{0.5cm}
Fig.\ref{MDFIG}
shows the evolution of the logarithmic distance, $\ln D(t),$
for an $N=6$ cubic lattice with the total scaled energy
per link of $g^2aE=2.00$. The different curves counted from above
correspond to charges $q=12.10,$ $0.0,$ $8.70,$ $9.50$
at the ends of the flux line and to
respectively reduced random magnetic backgrounds, belonging to
an independent random choice of SU(2) group elements with the
main group angle covering uniformly $45\%,$ $100\%,$ $35\%$ and
$15\%$ of the interval $[0,2\pi]$.

\vspace{0.5cm}
A tendency of faster initial oscillations in the divergence
of the two parting configurations with increasing strength
of the flux line can be observed. It is followed
by linear epoches with a reduced slope, i.e. reduced
chaotic dynamics.
Correspondingly the entropy generation is also reduced.

\vspace{0.5cm}
The equipartition of the chromoelectric and chromomagnetic
energy is essentially faster than the entropy generation rate
set by the leading Lyapunov exponent. Of course the partition
of the total lattice energy (not counting the energy of the static
charge) develops already in the linear approximation without
any chaotic behavior. Its time scale is determined by the
highest oscillation frequency, $\sim \pi/a,$ of the lattice system
and after a few oscillation periods the equipartition is completed.

\vspace{0.5cm}
\begin{figure}[p]
\vspace{0.3cm}
\centerline{\rotate[r]{\psfig{figure=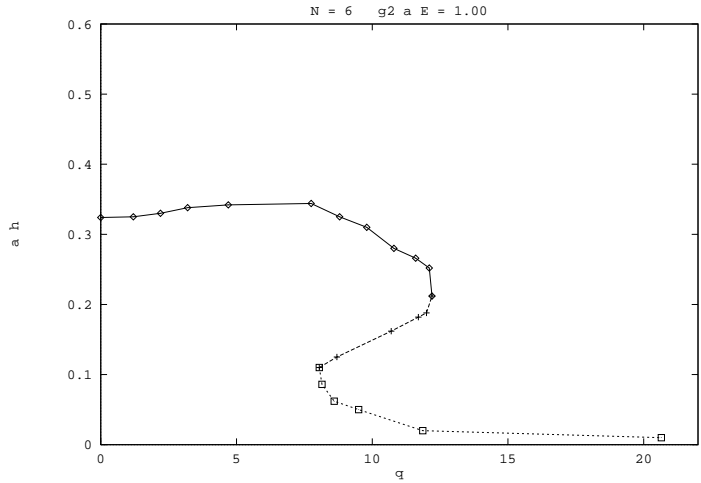,width=6cm}}}
\vspace{0.3cm}
\caption{\label{CHFIG} The leading Lyapunov exponent as function of the static
charges positioned at the ends of a diagonal flux line.}
\end{figure}
\vspace{0.5cm}

\vspace{0.5cm}
A systematic study of the leading Lyapunov exponents as a
function of the charge strength at the ends of the initial flux line
reveals an $S$-like structure known from first order phase
transitions (cf. Fig.\ref{CHFIG}).
The slight initial rise of the upper branch may be due to
numerical uncertainties; the $N=6$ system is relatively small
in linear size, although the allowed phase space is 3239 - dimensional.
The qualitative feature of a microcanonical simulation of a
phase transition between weak and strong charges is, however,
inspiring.

\vspace{1.5cm}
{\em In conclusion} these simulations have shown that the
chaotic dynamics of classical SU(2) Yang-Mills fields
survives in the presence of weak static charges and a thin flux
line of chromoelectric energy spanned between them.
In order to reduce the chaos essentially a strong
(in the order of magnitude of 10 in scaled units) charge
is necessary.
This reduction occurs then suddenly at a given critical strength
and shows a hysteresis -- like in case of a ferromagnetic
phase transition.

\vspace{0.5cm}
With respect to relativistic heavy ion collisions this result
leaves room for the hope that chromoelectric strings (color ropes)
formed initially share their energy with random chromomagnetic fields
-- which bundle them together -- and therefore the soft collective
chaotic dynamics of these fields is not suppressed
essentially. This chaotic dynamics then contributes to the thermalization
of primordial quark matter helping it to evolve towards
a quark - gluon plasma.


\vspace{1.5cm}
{\bf Acknowledgements}
\vspace{0.5cm}

This work was supported by the Deutsche Forschungsgemeinschaft
and the Hungarian Academy of Science (DFG-MTA 79/1994).
The warm hospitality of the Institute for Theoretical Physics
of University of Giessen, where an essential part
of this work has been done,  is gratefully acknowledged.



\end{document}